\documentclass[aps,prl,reprint,showpacs,floatfix,twocolumn]{revtex4-1}
\usepackage{graphicx}
\usepackage{amsmath}
\usepackage{nicefrac}
\usepackage[pdftex]{hyperref}

\begin{document}

\title{Quantum interference in the absorption and emission\\ of single photons by a single ion}

\author{M. Schug}
\email{mschug@physik.uni-saarland.de}
\author{C. Kurz}
\author{P. Eich}
\author{J. Huwer}
\author{P. M\"uller}
\author{J. Eschner}

\affiliation{
Universit\"at des Saarlandes, Experimentalphysik, Campus E2 6, 66123 Saarbr\"ucken, Germany\\
}

\date{\today}

\begin{abstract}
We investigate quantum beats in the arrival-time distribution of single photons from a single trapped $^{40}$Ca$^+$ ion, revealing their fundamentally different physical origins in two distinct experimental situations: In a $\Lambda$-type level scheme the interference of two 854-nm \emph{absorption} amplitudes suppresses and enhances the emission process of Raman-scattered 393-nm photons; in a V-type level scheme the interference of two 393-nm \emph{emission} amplitudes causes a rotation of their dipole emission pattern resulting in a temporal modulation of the detected photons. For both cases we demonstrate coherent control over the quantum-beat phase through the phases of the atomic and photonic input states, which also allows controlled adjustment of the total photon detection efficiency.
\end{abstract}


\maketitle

\section{I. Introduction}
Since the early description of quantum mechanics, the effect of quantum interference led to fascinating and peculiar predictions, one being the appearance of quantum beats caused by interference of emission paths from excited atoms \cite{Breit1933}. 
Nowadays, quantum interference is an essential feature in a quantum network based on quantum optical systems with single atoms as nodes and single photons transferring information between them \cite{Cirac1997,Kimble2008}. It allows for entanglement between two nodes \cite{Moehring2007,Slodicka2013,Hofmann2012} and employing coherent phenomena at atom-photon interfaces to faithfully convert quantum information between photonic communication channels and atomic quantum processors \cite{Ritter2012}. In this context, the controlled emission \cite{Blinov2004,Maunz2007,Almendros2009,Kurz2013} and absorption \cite{Piro2011,Huwer2013} of single photons by a single atom is crucial for schemes proposing the heralded mapping of a photonic polarization state into an atomic quantum memory \cite{Muller2013}. Here we show the quantum coherent character of the absorption and emission of single photons through the interference between indistinguishable quantum channels in a single atom.
 
First experimental observations of quantum beats were induced by pulsed optical excitation of atoms with two upper states which decay to the same ground state \cite{Dodd1967, Aleksandrov1964}. In continuously excited atomic ensembles, the observation of quantum beats in the correlation of two photons was demonstrated for a coherent superposition state with short lifetime in a calcium cascade \cite{Aspect1984} and in cavity mediated systems with coherent ground states \cite{Norris2010}. For a single ion \cite{Schubert1995}, transient effects of the internal dynamics showed photonic oscillations by interference in absorption. 

Here we generate controlled, high-contrast quantum beats in a spontaneous Raman scattering process which consists of the emission of single 393-nm photons after absorption of 854-nm laser photons in a single trapped $^{40}\text{Ca}^+$ ion. We consider two distinct excitation schemes, called $\Lambda$ and V. For both schemes we employ the controlled generation of an initial coherent superposition state of two Zeeman sublevels in the metastable D$_{5/2}$ manifold with phase-coherent laser pulses. Quantum beats are then observed in the arrival-time distribution of the detected 393-nm photons after onset of the 854-nm laser pulse that induces their emission. We show that the observed quantum beats are fully controlled through changes of the phase in the atomic superposition and the photonic polarization input states. In particular, we highlight two distinct physical origins of the quantum beats, namely, quantum interference of two 854-nm absorption amplitudes and two 393-nm emission amplitudes, respectively. The two absorption-emission pathways resemble a which-way experiment where indistinguishability is afforded by a quantum eraser \cite{Scully1982}.    

\section{II. Experimental sequence}

\begin{figure}[htbp]
\centering
\includegraphics[width=8.6cm]{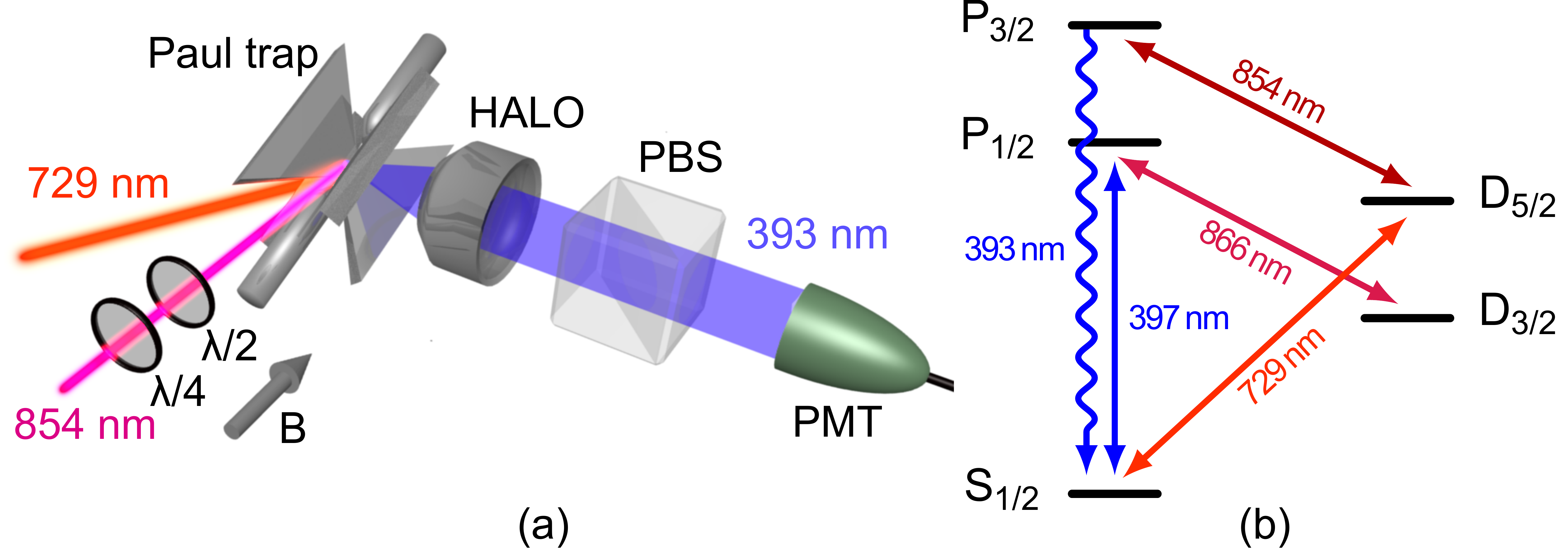}
\caption{(Color online) (a) Schematic of the experimental setup: excitation of a single trapped ion with an 854-nm laser beam parallel to the magnetic field axis and polarization-selective detection of 393-nm photons perpendicular to it. HALO stands for high-numerical-aperture laser objective \cite{Gerber2009}, PBS for polarizing beam splitter, and PMT for photomultiplier tube. (b) Level scheme and relevant transitions of the $^{40}$Ca$^+$ ion.}
\label{setup} 
\end{figure}

The experimental setup is illustrated in Fig.~\ref{setup}(a). A single $^{40}$Ca$^+$ ion is trapped in a linear Paul trap and excited by various laser beams. A magnetic field $B$ parallel to the trap axis defines the quantization axis and lifts the degeneracy of the atomic levels [Fig.~\ref{setup}(b)]. The experimental sequence starts with Doppler cooling on the S$_{1/2}$-P$_{1/2}$ transition with a 397-nm laser, aided by an 866-nm laser that repumps the population from the metastable D$_{3/2}$ state. After cooling, a circularly polarized laser pulse optically pumps the ion to $\left\lvert\rm{S}_{1/2},-\frac{1}{2}\right\rangle$. Then it is excited to two selected Zeeman sublevels in the metastable D$_{5/2}$ manifold, employing a laser at 729~nm which is locked to an ultra-stable high-finesse cavity. Efficient population transfer up to 99.6(3)$\%$ on each of the two transitions is achieved by frequency-selective coherent pulses. From the difference of the two frequencies we extract directly the Larmor frequency $\nu_{L}$ in D$_{5/2}$ [see Eq.~(\ref{Zeeman-splitting}) below]. The superposition state is excited to the $\text{P}_{3/2}$ level by a laser at 854-nm wavelength with controlled polarization and frequency. Its detuning $\Delta$ from the line center is previously calibrated by a spectroscopic measurement and it is adjusted to provide identical pumping rates for the two initial Zeeman sublevels. The 854-nm beam enters parallel to the quantization axis and its polarization is adjustable to any defined linear or circular state by a combination of $\lambda/4$ and $\lambda/2$ waveplates. The absorption of the 854-nm laser photons leads to the emission of a 393-nm Raman-scattered photon that is collected perpendicular to the quantization axis by an in-vacuum high-numerical-aperture laser objective (HALO) with a numerical aperture of 0.4 \cite{Gerber2009}. The photons are selected by their polarization with a polarizing beam splitter (PBS) cube and detected by a photomultiplier tube (PMT). The PMT pulses are fed into a time-correlated single-photon counting module (Pico Harp 300) for temporally correlating them with the sequence trigger, i.e., with the onset of the 854-nm pulse. 

An inherent requirement for quantum interference between two scattering paths is to keep indistinguishability in all degrees of freedom of the involved quantum channels. In the $\Lambda$- and V-type level configurations, the two absorption channels exhibit a frequency difference originating from the differential Zeeman shift of the two D$_{5/2}$ sublevels, up to $\sim$20~MHz for typical magnetic fields of $\sim$3~G. The PMT time resolution of 300~ps sets a much lower frequency resolution and thus erases the information of this frequency splitting \cite{Togan2010}. The indistinguishability concerning the polarization is attained by the detection perpendicular to the magnetic field of only $\left\lvert\rm{H}\right\rangle$-polarized or only $\left\lvert\rm{V}\right\rangle$-polarized 393-nm photons (i.e.,\ polarization parallel or orthogonal to the quantization axis, respectively).

\section{III. Theoretical analysis} 

A brief and simplified theoretical description of the population transfer from D$_{5/2}$ to S$_{1/2}$ is presented which emphasizes the coherent evolution of the internal states for the two different level configurations.

\begin{figure}[htbp]
\centering
\includegraphics[width=8.4cm]{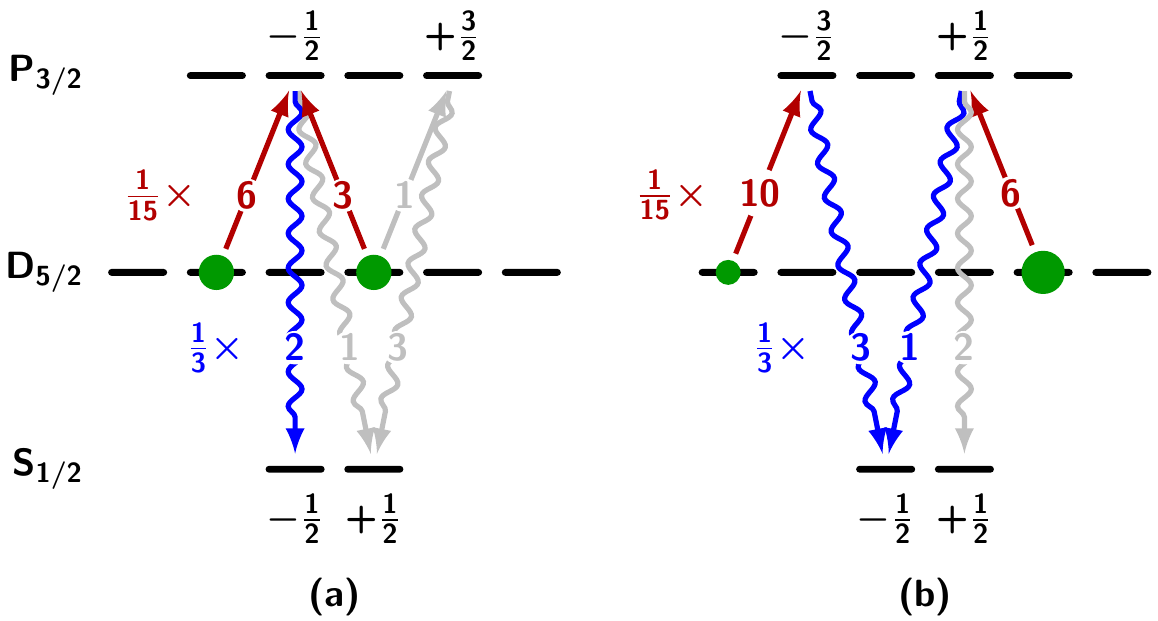}
\caption{(Color online) (a) The $\Lambda$-shaped three-level system consisting of
$\left\lvert\rm{D}_{5/2},-\frac{3}{2}\right\rangle$, $\left\lvert\rm{D}_{5/2},+\frac{1}{2}\right\rangle$ and $\left\lvert\rm{P}_{3/2},-\frac{1}{2}\right\rangle$ including the relevant transitions with their squared Clebsch-Gordan coefficients. (b) The V-shaped three-level system consisting of
$\left\lvert\rm{P}_{3/2},-\frac{3}{2}\right\rangle$, $\left\lvert\rm{P}_{3/2},+\frac{1}{2}\right\rangle$ and $\left\lvert\rm{S}_{1/2},-\frac{1}{2}\right\rangle$ including the relevant transitions with their squared Clebsch-Gordan coefficients. Red arrows, absorption of 854-nm photons ($\sigma^+$ and $\sigma^-$); blue wavy arrows, emission of 393-nm photons; gray arrows, parasitic absorption (854~nm) and emission channels (393~nm).}
\label{levelscheme_both}
\end{figure}

Figure~\ref{levelscheme_both} shows the $\Lambda$- and V-type level configurations with the relevant transitions and the squared Clebsch-Gordan coefficients (CGCs).

Both schemes are initialized in a coherent superposition state,  
\begin{equation}
\left\lvert\psi_{\rm{D}}(t)\right\rangle = \sqrt{\rho_1} \left\lvert \text{D}_{5/2},m_\text{D}\right\rangle + \text{e}^{i\Phi_{\text{D}}(t)} \sqrt{\rho_2}\left\lvert \text{D}_{5/2},m_{\text{D'}}\right\rangle, 
\label{initial_state}
\end{equation}
with populations $\rho_1$ and $\rho_2$ adjusted by two consecutive Rabi pulses from $\left\lvert\rm{S}_{1/2},-\frac{1}{2}\right\rangle$. The phase $\Phi_{\rm{D}}(t)=\Phi_{\rm{D}}(0)+2\pi\nu_{L}t$ is composed of a starting phase $\Phi_{\rm{D}}(0)$ which is set by the sequence control as the relative phase between the two 729-nm pulses and a part oscillating with the Larmor frequency,
\begin{equation}
 \nu_{L}=\frac{\mu_B}{h}\Delta m_j \,g_j\,B,
 \label{Zeeman-splitting}
\end{equation} 
including the magnetic field $B$, the Land$\acute{\rm{e}}$ factor $g_j=\frac{6}{5}$, and the Bohr magneton $\mu_B$.

The polarization state of the laser photons at 854~nm is defined as a superposition of two orthogonal states, namely, right ($\left\lvert\rm{R}\right\rangle$) and left ($\left\lvert\rm{L}\right\rangle$) circularly polarized light 
\begin{equation}
\left\lvert\psi_{854}\right\rangle=\rm{cos}\tfrac{\vartheta}{2}\left\lvert\rm{R}\right\rangle+\rm{sin}\tfrac{\vartheta}{2}\rm{e}^{i\Phi_{854}}\left\lvert\rm{L}\right\rangle.
\label{state854}
\end{equation}
Any linear photonic polarization state, such as horizontal $\left\lvert\rm{H}\right\rangle$, vertical $\left\lvert\rm{V}\right\rangle$, diagonal $\left\lvert\rm{D}\right\rangle$, and antidiagonal $\left\lvert\rm{A}\right\rangle$ is adjusted by rotating the two wave plates [see Fig.~\ref{setup}(a)] to change $\Phi_{854}$, while $\vartheta$ is fixed to $\pi/2$. As the propagation direction of the incoming laser photons is chosen parallel to the applied magnetic field, the photon polarization translates to the reference frame of the atom according to \begin{equation}
\left\lvert\psi_{854}\right\rangle = \rm{cos}\tfrac{\vartheta}{2}\left\lvert+1\right\rangle + \rm{sin}\tfrac{\vartheta}{2}\rm{e}^{i\Phi_{854}}\left\lvert-1\right\rangle,
\end{equation}
whereby $\left\lvert m_{854} \right\rangle = \left\lvert \pm1 \right\rangle$ stand for the photon polarizations that effect $\Delta m = \pm1$ (i.e., $\sigma^{\pm}$) transitions, respectively, between the Zeeman sublevels of D$_{5/2}$ and P$_{3/2}$. 

The coupled quantum system is represented by a joint state between photon and atom,
\begin{equation}
\left\vert\psi(t)\right\rangle=\left\lvert\psi_{\rm{D}}(t)\right\rangle\otimes \left\lvert\psi_{854}\right\rangle. 
\label{joint}
\end{equation}
Based on \cite{Muller2013} the absorption process is described by
\begin{equation}
\hat{A}=\sum_{m_{\rm{D}},m_{\rm{P}}}C_{m_{\rm{D}},m_{854},m_{\rm{P}}} ~ \tilde{c}_{m_{\rm{D}},m_{\rm{P}}}(\Delta) \left\lvert m_{\rm{P}}\right\rangle\left\langle m_{\rm{D}}\right\lvert\left\langle m_{854}\right\lvert
\label{absop}
\end{equation}
where $m_{854}=0,\pm1$, and $C_{m_{\rm{D}},m_{854},m_{\rm{P}}}$ are the CGCs \cite{footnoteCGC}. The detuning-dependent atomic response $\tilde{c}_{m_{\rm{D}},m_{\rm{P}}}(\Delta)=\lvert \tilde{c}(\Delta)\rvert\rm{e}^{i\phi(\Delta)}$ has a complex Lorentzian lineshape with linewidth $\Gamma_{\rm{P}_{3/2}}$. We combine these coefficients according to 
\begin{equation}
C_{m_{\rm{D}},m_{854},m_{\rm{P}}} ~ \tilde{c}_{m_{\rm{D}},m_{\rm{P}}}(\Delta) = c_{m_{\rm{D}},m_{\rm{P}}}(\Delta)
\end{equation}
using $m_{\rm{P}}=m_{\rm{D}}+m_{854}$. The detuning $\Delta=\omega_{\rm{L}}-\omega_{0}$ between the laser frequency $\omega_{\rm{L}}$ and the D$_{5/2}$ to P$_{3/2}$ line center $\omega_{0}$ will be used as a control knob to compensate for the different CGCs in the absorption channels.

After the absorption, the ion decays in a spontaneous (Raman) emission process to the S$_{1/2}$ ground state, described by an emission operator,
\begin{equation}
\hat{E}=\sum_{m_{\rm{S}},m_{\rm{P}}}C_{m_{\rm{P}},m_{393},m_{\rm{S}}}\left\lvert m_{393}\right\rangle\left\lvert m_{\rm{S}}\right\rangle\left\langle m_{\rm{P}}\right\lvert,
\label{emop}
\end{equation}
with $m_{393}=0,\pm1$. Applying the absorption and emission operator to the joint state of Eq.~(\ref{joint}) gives $\hat{E}\hat{A}\left\vert\psi(t)\right\rangle$, a new joint state between an atom in S$_{1/2}$ and a single photon in the 393-nm mode, which is projected onto $\left\lvert \rm{S}_{1/2},-\tfrac{1}{2}\right\rangle$ conditioned on the detection of a 393-nm photon with certain polarization. 

\subsection{A. The \texorpdfstring{$\Lambda$}{Lambda} system}

In the case of the $\Lambda$ system, the ion is prepared in the coherent superposition state 
\begin{equation}
\left\lvert\psi_{\rm{D}}(t)\right\rangle=\sqrt{\tfrac{1}{2}}\left(\left\lvert\rm{D}_{5/2},-\tfrac{3}{2}\right\rangle+\rm{e}^{i\Phi_{\rm{D}}(t)}\left\lvert\rm{D}_{5/2},+\tfrac{1}{2}\right\rangle\right),
\label{L-superposition}
\end{equation}
such that two absorption paths share the same emission channel [Fig.~\ref{levelscheme_both}(a)]. Applying the absorption operator to the joint state gives
\begin{equation}
\begin{split}
\hat{A}\left\lvert\psi(t)\right\rangle &= \sqrt{\tfrac{1}{2}}\cos\tfrac{\vartheta}{2}~c_{-\nicefrac{3}{2},-\nicefrac{1}{2}}(\Delta) \left\lvert\rm{P}_{3/2},-\tfrac{1}{2}\right\rangle\\
&+\sqrt{\tfrac{1}{2}}\sin\tfrac{\vartheta}{2}~c_{+\nicefrac{1}{2},-\nicefrac{1}{2}}(\Delta)~\rm{e}^{i\Phi_{\rm{D}}(t)}\rm{e}^{i\Phi_{854}}\left\lvert\rm{P}_{3/2},-\tfrac{1}{2}\right\rangle\\
&+\sqrt{\tfrac{1}{2}}\cos\tfrac{\vartheta}{2}~c_{+\nicefrac{1}{2},+\nicefrac{3}{2}}(\Delta)~\rm{e}^{i\Phi_{\rm{D}}(t)}\left\lvert\rm{P}_{3/2},+\tfrac{3}{2}\right\rangle.
\end{split}
\end{equation}
Highest visibility of the quantum beats is expected when the interfering transitions have equal weights. The amplitudes of the two absorbing paths are determined by the detuning $\Delta$, which is adjusted in order to compensate for the two different CGCs, i.e., such that $|c_{-\nicefrac{3}{2},-\nicefrac{1}{2}}(\Delta)| = |c_{+\nicefrac{1}{2},-\nicefrac{1}{2}}(\Delta)| = c$.  
For a linear photonic polarization state ($\vartheta=\tfrac{\pi}{2}$) it follows that 
\begin{equation}
\begin{split}
\hat{A}\left\lvert\psi(t)\right\rangle &= \tfrac{1}{2}c\left(1+\rm{e}^{i\Phi_{\rm{D}}(t)}\rm{e}^{i\Phi_{854}}\right)\left\lvert\rm{P}_{3/2},-\tfrac{1}{2}\right\rangle\\
& \quad
+\tfrac{1}{2}c'\rm{e}^{i\Phi_{\rm{D}}(t)}\left\lvert\rm{P}_{3/2},+\tfrac{3}{2}\right\rangle.
\end{split}
\label{absorbH}
\end{equation}
with $c' = c_{+\nicefrac{1}{2},+\nicefrac{3}{2}}(\Delta)$. Here the term $\left(1+\rm{e}^{i\Phi_{\rm{D}}(t)}\rm{e}^{i\Phi_{854}}\right)$ already shows the interference of the two absorption paths in the amplitude of $\left\lvert\rm{P}_{3/2},-\frac{1}{2}\right\rangle$.
The transfer of this oscillation to the emitted 393-nm photons is obtained by applying the emission operator $\hat{E}$,
\begin{equation}
\begin{split}
\hat{E}\hat{A}\left\lvert\psi(t)\right\rangle &= \sqrt{\tfrac{1}{6}}c\left(1+\rm{e}^{i\Phi_{\rm{D}}(t)}\rm{e}^{i\Phi_{854}}\right)\left\lvert 0 \right\rangle\left\lvert\rm{S}_{1/2},-\tfrac{1}{2}\right\rangle\\
&+\sqrt{\tfrac{1}{12}}c\left(1+\rm{e}^{i\Phi_{\rm{D}}(t)}\rm{e}^{i\Phi_{854}}\right)\left\lvert -1 \right\rangle\left\lvert\rm{S}_{1/2},+\tfrac{1}{2}\right\rangle\\
&+\sqrt{\tfrac{1}{8}}c'\rm{e}^{i\Phi_{\rm{D}}(t)}\left\lvert +1 \right\rangle\left\lvert\rm{S}_{1/2},+\tfrac{1}{2}\right\rangle.
\end{split}
\label{state1}
\end{equation}
Of these three different decay channels, selection of $\Delta m = 0$, i.e., of $\pi$ photons, is favorable for keeping indistinguishability between the two interfering scattering channels. A $\pi$ photon detected perpendicular to the magnetic field transforms to an $\left\lvert\rm{H}\right\rangle$-polarized photon in the photonic reference frame. The detection of these photons projects the joint state of Eq.~(\ref{state1}) onto $\left\lvert\rm{S}_{1/2},-\tfrac{1}{2}\right\rangle$. The intensity of the emitted light is proportional to 
\begin{equation}
I \propto \tfrac{1}{3} c^2(1+\text{cos}\,(\Phi_{\text{D}}(t)+\Phi_{854}))
\label{intensity1}
\end{equation}
and shows the possibility to be controlled by changing the photonic input phase $\Phi_{854}$ or the atomic superposition phase $\Phi_{\rm{D}}(t)$ through the offset phase $\Phi_{\rm{D}}(0)$. We note again that interference happens in the absorption process, since two pathways lead to the same excited intermediate state before the emission process takes place.

\subsection{B. The V system}
The V-type configuration and the corresponding excitation scheme for the generation of 393-nm photons is shown in Fig.~\ref{levelscheme_both}(b). Starting in the eigenstate $\left\lvert\rm{S}_{1/2},-\frac{1}{2}\right\rangle$, the first resonant 729-nm pulse transfers 75$\%$ of the population to $\left\lvert\rm{D}_{5/2},+\frac{3}{2}\right\rangle$. The remaining population is subsequently transferred to $\left\lvert\rm{D}_{5/2},-\frac{5}{2}\right\rangle$  with a $\pi$ pulse, resulting in the coherent superposition state
\begin{equation}
\left\lvert\psi_{\rm{D}}(t)\right\rangle=\sqrt{\tfrac{1}{4}}\left\lvert\rm{D}_{5/2},-\tfrac{5}{2}\right\rangle+\rm{e}^{i\Phi_{\rm{D}}(t)}\sqrt{\tfrac{3}{4}}\left\lvert\rm{D}_{5/2},+\tfrac{3}{2}\right\rangle. 
\label{V-superposition}
\end{equation}
The unequal squared CGCs of the two 393-nm decay channels, i.e., the spurious decay into unwanted channels [see Fig.~\ref{levelscheme_both}(b)], is compensated for by the unequal initial population distribution in D$_{5/2}$. Applying the absorption operator (\ref{absop}) to the joint state gives
\begin{equation}
 \begin{split}
\hat{A}\left\lvert\psi(t)\right\rangle
 &=\sqrt{\tfrac{1}{4}}\cos\tfrac{\vartheta}{2}~c_{-\nicefrac{5}{2},-\nicefrac{3}{2}}(\Delta)\left\lvert\rm{P}_{3/2},-\tfrac{3}{2}\right\rangle\\
& +\sqrt{\tfrac{3}{4}}\sin\tfrac{\vartheta}{2}~c_{+\nicefrac{3}{2},+\nicefrac{1}{2}}(\Delta)~\rm{e}^{i\Phi_{\rm{D}}(t)} \rm{e}^{i\Phi_{854}}\left\lvert\rm{P}_{3/2},+\tfrac{1}{2}\right\rangle.
 \end{split}
\end{equation}
The two different CGCs on the 854-nm absorption channels are again compensated by two different 854-nm repumping rates, set by adjusting the detuning $\Delta$ such that $|c_{-\nicefrac{5}{2},-\nicefrac{3}{2}}(\Delta)| = |c_{+\nicefrac{3}{2},+\nicefrac{1}{2}}(\Delta)| = c$. For a linear 854-nm polarization with phase $\Phi_{854}$ we get 
\begin{equation}
 \begin{split}
\hat{A}\left\lvert\psi(t)\right\rangle&=\sqrt{\tfrac{1}{8}}c\left\lvert\rm{P}_{3/2},-\tfrac{3}{2}\right\rangle+\sqrt{\tfrac{3}{8}}c~\rm{e}^{i\Phi_{\rm{D}}(t)}\rm{e}^{i\Phi_{854}}\left\lvert\rm{P}_{3/2},+\tfrac{1}{2}\right\rangle.\\
 \end{split}
\end{equation}

Applying the emission operator (\ref{emop}) results in
\begin{equation}
 \begin{split}
\hat{E}\hat{A}\left\lvert\psi(t)\right\rangle &= \left(\left\lvert -1 \right\rangle+\rm{e}^{i\Phi_{\rm{D}}(t)}\rm{e}^{i\Phi_{854}}\left\lvert +1 \right\rangle\right)\sqrt{\tfrac{1}{8}}c\left\lvert\rm{S}_{1/2},-\tfrac{1}{2}\right\rangle\\
&+ \tfrac{1}{2}c~\rm{e}^{i\Phi_{\rm{D}}(t)}\rm{e}^{i\Phi_{854}}\left\lvert 0 \right\rangle\left\lvert\rm{S}_{1/2},+\tfrac{1}{2}\right\rangle.
\label{superpos}
 \end{split}
\end{equation}
The first term in this state shows the oscillating phase between two atomic ($\sigma^{\pm}$) transitions leading to interference of two emission channels when they are projected on the same axis. More precisely, the superposition of $\sigma^-$ and $\sigma^+$ emission causes the dipolar emission pattern to rotate with the Larmor frequency about the quantization axis, which leads to a temporal modulation of the detected photons in the direction perpendicular to that axis. Measurement of the photonic superposition $\left\lvert\rm{V}\right\rangle=\sqrt{\frac{1}{2}}\left(\left\lvert\sigma^+\right\rangle-\left\lvert\sigma^-\right\rangle\right)$ projects the joint state of Eq.~(\ref{superpos}) onto $\left\lvert\rm{S}_{1/2},-\tfrac{1}{2}\right\rangle$, which results in an intensity  
\begin{equation}
I \propto \tfrac{1}{8}c^2(1-\rm{cos}\,(\Phi_{\rm{D}}(t)+\Phi_{854})),
\label{intensity}
\end{equation}
showing again oscillations with the Larmor frequency.
The second term in (\ref{superpos}) describes the emission of parasitic $\pi$ photons which transforms in the photonic reference frame to $\left\lvert\rm{H}\right\rangle$. They are suppressed by rotating the PBS by $90^{\circ}$ with respect to the $\Lambda$ case.

\section{IV. Experimental Results}
\subsection{A. Quantum beats in arrival-time distributions}

\begin{figure}[htbp]
\centering
\includegraphics[width=8.4cm]{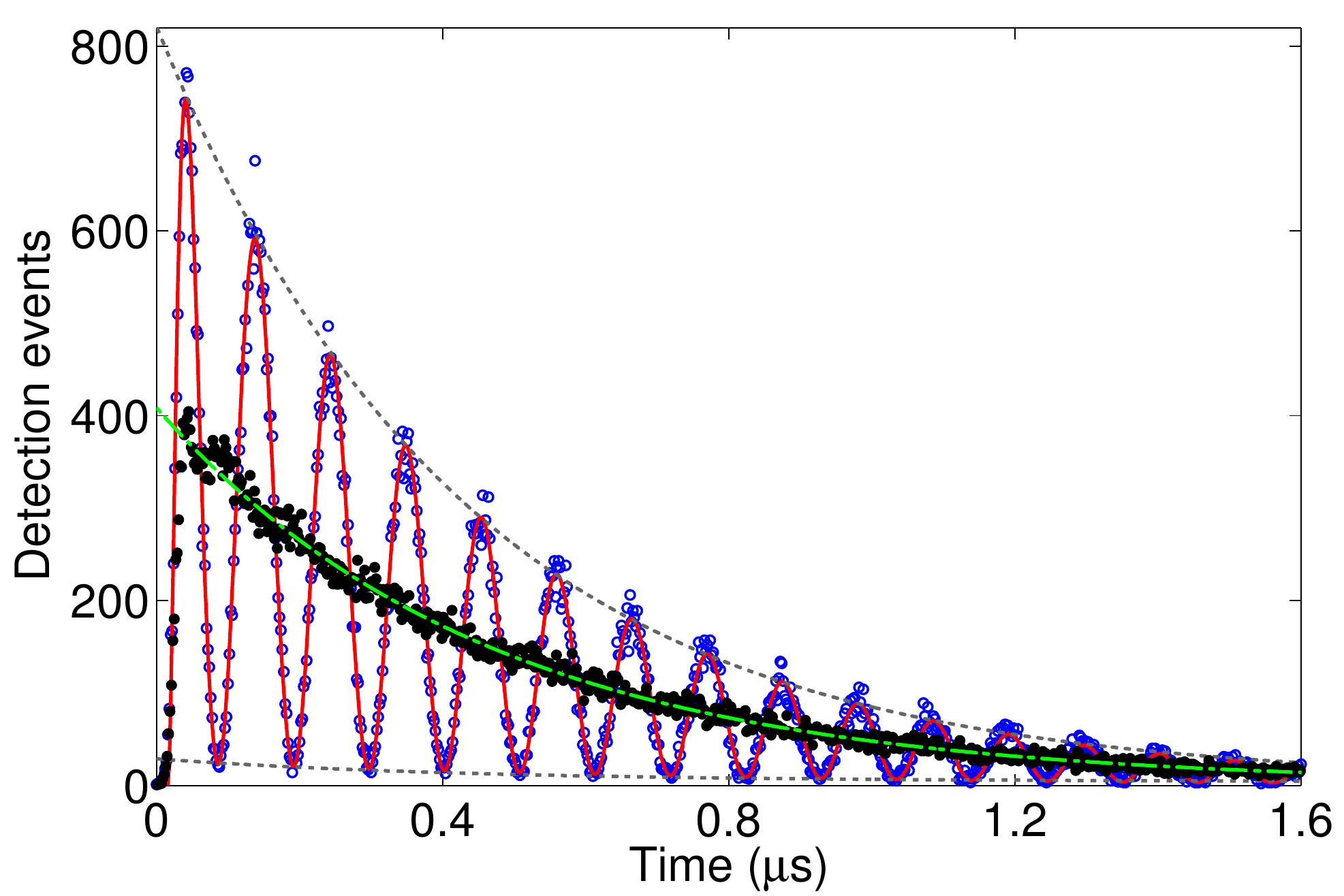}
\caption{(Color online) Arrival-time distribution of single 393 nm photons from an initial coherent superposition (blue circles) and statistical mixture (black dots) in D$_{5/2}$. Red solid line, fit to the data by numerically solving the 18-level optical Bloch equations; green dash-dotted line, exponential fit to the data for the mixture; gray dotted line, exponential fit to the envelope of the oscillation, from which the visibility is determined.}
\label{L-incoherent}
\end{figure}

We first discuss the experimental results for the $\Lambda$ scheme. The blue circles in Fig.~$\ref{L-incoherent}$ show the arrival-time distribution of the 393-nm Raman-scattered photons for the case of an initial coherent superposition between $\left\lvert\rm{D}_{5/2},-\frac{3}{2}\right\rangle$ and $\left\lvert\rm{D}_{5/2},+\frac{1}{2}\right\rangle$. The exponential decay of the photon wave packet is modulated with a period of 106 ns, corresponding to a Larmor frequency in D$_{5/2}$ of 9.4 MHz, which is in agreement with the frequency difference of the initially populated Zeeman sublevels, as determined from 729-nm spectroscopy. The data points are fitted by numerically solving the optical Bloch equations, including all relevant Zeeman sublevels and the projection of the final joint state [Eq.~(\ref{state1})] according to the detection of an $\left\lvert\rm{H}\right\rangle$-polarized photon. The visibility of the oscillation, determined from the envelope of the fit (gray dotted line), is 93.1(6)$\%$. For comparison, we also show the photon arrival-time distribution for the case of an initial statistical mixture in D$_{5/2}$ (black dots). The data set is generated by averaging the two individually recorded arrival-time distributions for the $\left\lvert\rm{D}_{5/2},-\frac{3}{2}\right\rangle$ to $\left\lvert\rm{S}_{1/2},-\frac{1}{2}\right\rangle$ and  $\left\lvert\rm{D}_{5/2},+\frac{1}{2}\right\rangle$ to $\left\lvert\rm{S}_{1/2},-\frac{1}{2}\right\rangle$ scattering processes. From the fit of the exponential decay (green line) we get an 1/e decay time of 461(2) ns. 

\begin{figure}[htbp]
\centering
\includegraphics[width=8.3cm]{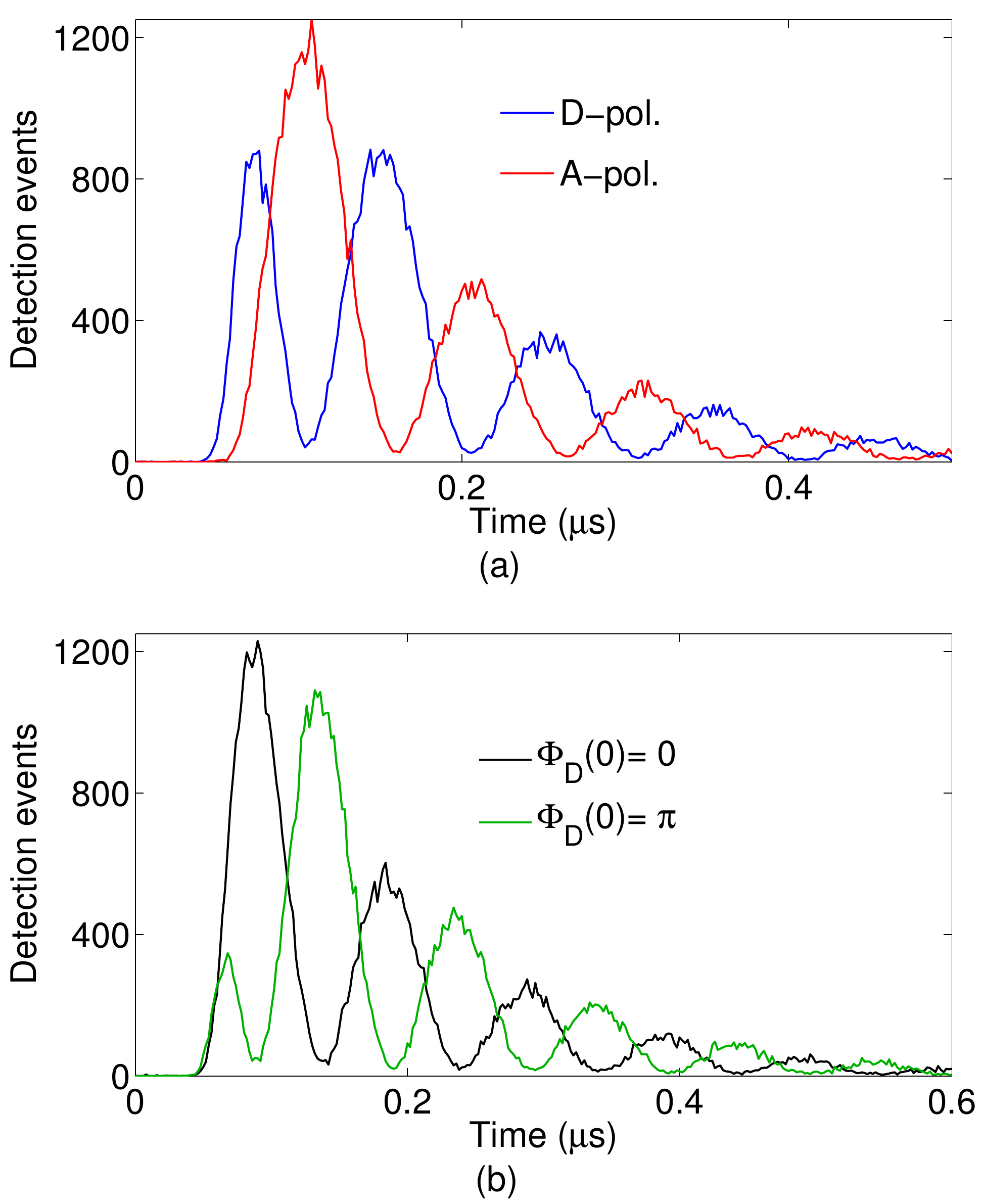}
\caption{(Color online) $\Lambda$ scheme. (a) Arrival-time distributions of the 393-nm photons showing quantum beats for two different 854-nm input polarization states, $\left\lvert\rm{D}\right\rangle$ and $\left\lvert\rm{A}\right\rangle$. (b) Arrival-time distributions showing quantum beats for two values, 0 and $\pi$, of the phase $\Phi_{\rm D}(0)$ of the initial atomic superposition. In (a) and (b), the bin size is 2~ns for 6~min measurement time. }
\label{L-osc-pol+phase}
\end{figure}

In the following we show that we have full control over the quantum phases that enter into the quantum beats of Fig.~\ref{L-incoherent}. First we study their dependence on the linear polarization of the absorbed 854-nm photons. Rotation of the wave plates allows adjustment of the incoming linear polarization, i.e., of the photonic phase $\Phi_{854}$ in Eq.~(\ref{intensity1}), to the canonical basis states $\left\lvert\rm{H}\right\rangle$, $\left\lvert\rm{V}\right\rangle$, $\left\lvert\rm{D}\right\rangle$, and $\left\lvert\rm{A}\right\rangle$ (and any value in between). Figure~\ref{L-osc-pol+phase}(a) shows arrival-time distributions for the polarization input states $\left\lvert\text{D}\right\rangle$ and $\left\lvert\text{A}\right\rangle$ (for the sake of clarity, we only show these two). The phase difference of $\Delta \Phi_{854}=180^{\circ}$ set by the waveplates is revealed in the two oscillations; the value derived from fitting the Bloch equations is $178.2(1.6)^{\circ}$. 

\begin{figure}[htbp]
\centering
\includegraphics[width=8.3cm]{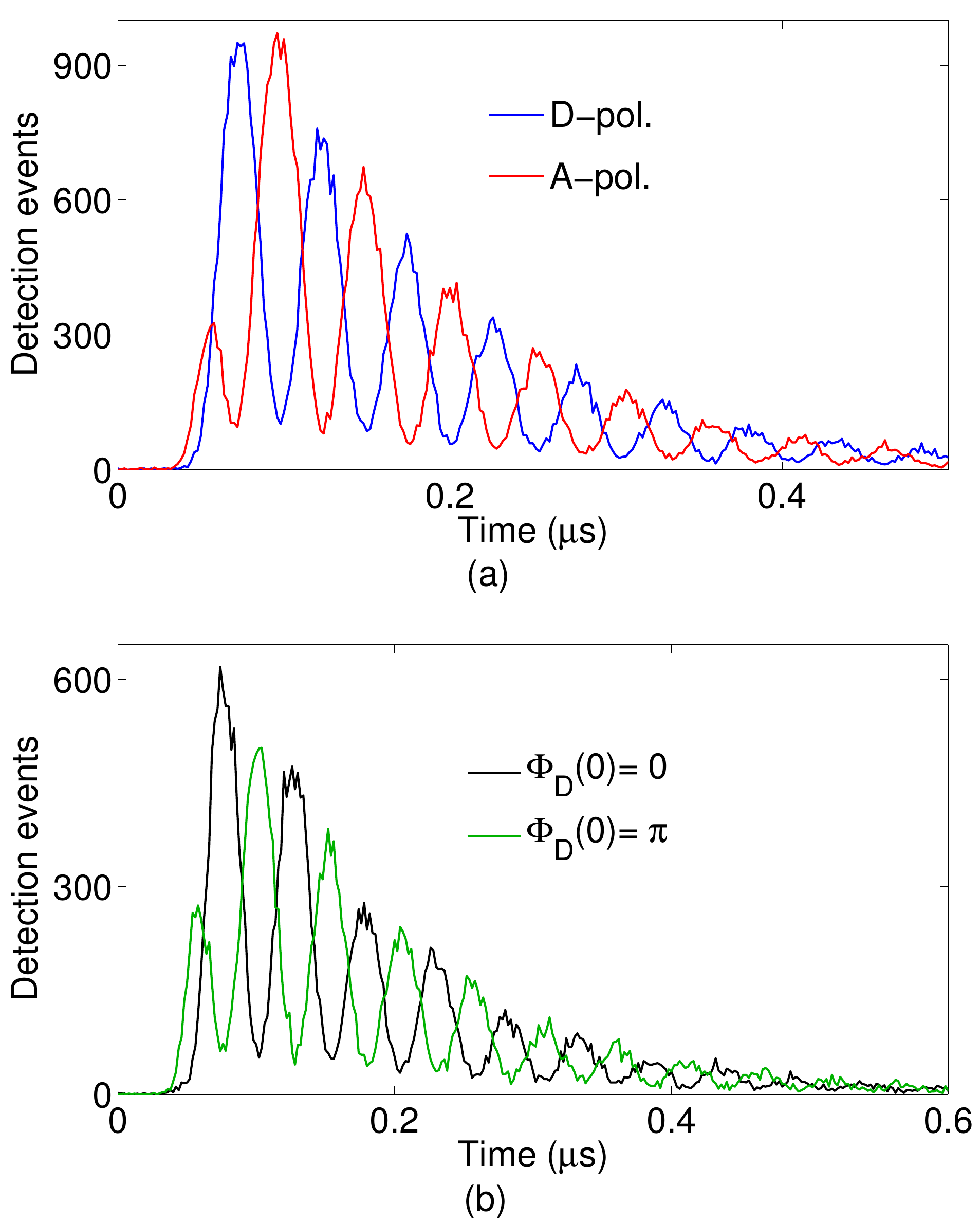}
\caption{(Color online) V scheme. (a) Quantum beats for two different 854-nm input polarization states, $\left\lvert\rm{D}\right\rangle$ and $\left\lvert\rm{A}\right\rangle$. (b) Quantum beats for two values, 0 and $\pi$, of the phase $\Phi_{\rm D}(0)$ of the initial atomic superposition. The bin size is 2~ns for 10~min measurement time [6~min in (b)]. }
\label{V-osc-pol+phase}
\end{figure}

Similarly, according to Eq.~(\ref{intensity1}), also the phase of the initial atomic superposition state $\Phi_{\rm D}(0)$ enters directly into the quantum beats. We control this phase via the radio-frequency source that drives the acousto-optic modulator setting the amplitude of the 729-nm laser. Figure~\ref{L-osc-pol+phase}(b) displays the change in the phase of the quantum beats effected by changing $\Phi_{\rm D}(0)$ by $180^{\circ}$, while keeping $\Phi_{854}$ constant; the Bloch equation fits yield a phase difference of $181.1(1.1)^{\circ}$. The $<1$\% deviation between the set values and the fitted values highlights the precise control that we have over the quantum beats. 
With the V scheme, we achieve analogous control through changes of both the photonic phase, i.e.\ the polarization of the incoming 854-nm photons, and the atomic phase, i.e.\ the difference phase of the preparing 729-nm pulses. Figure~\ref{V-osc-pol+phase}(a) displays the arrival-time distributions when the 854-nm polarization is adjusted to the orthogonal basis states $\left\lvert\rm{D}\right\rangle$ and $\left\lvert\rm{A}\right\rangle$. The Bloch equation fits yield a phase difference of 176.5(1.7)$^{\circ}$. In Fig.~\ref{V-osc-pol+phase}(b) the offset phase $\Phi_{\rm D}(0)$ of the coherent superposition in D$_{5/2}$ is set to 0 and $\pi$. Here the phase difference from the fits is $181.4(1.5)^{\circ}$. The deviation by 1-2\% from the ideal values confirms the precise control over the quantum-beat phase also for the case of the V scheme. 

\begin{figure}[htbp]
\centering
\includegraphics[width=8.3cm]{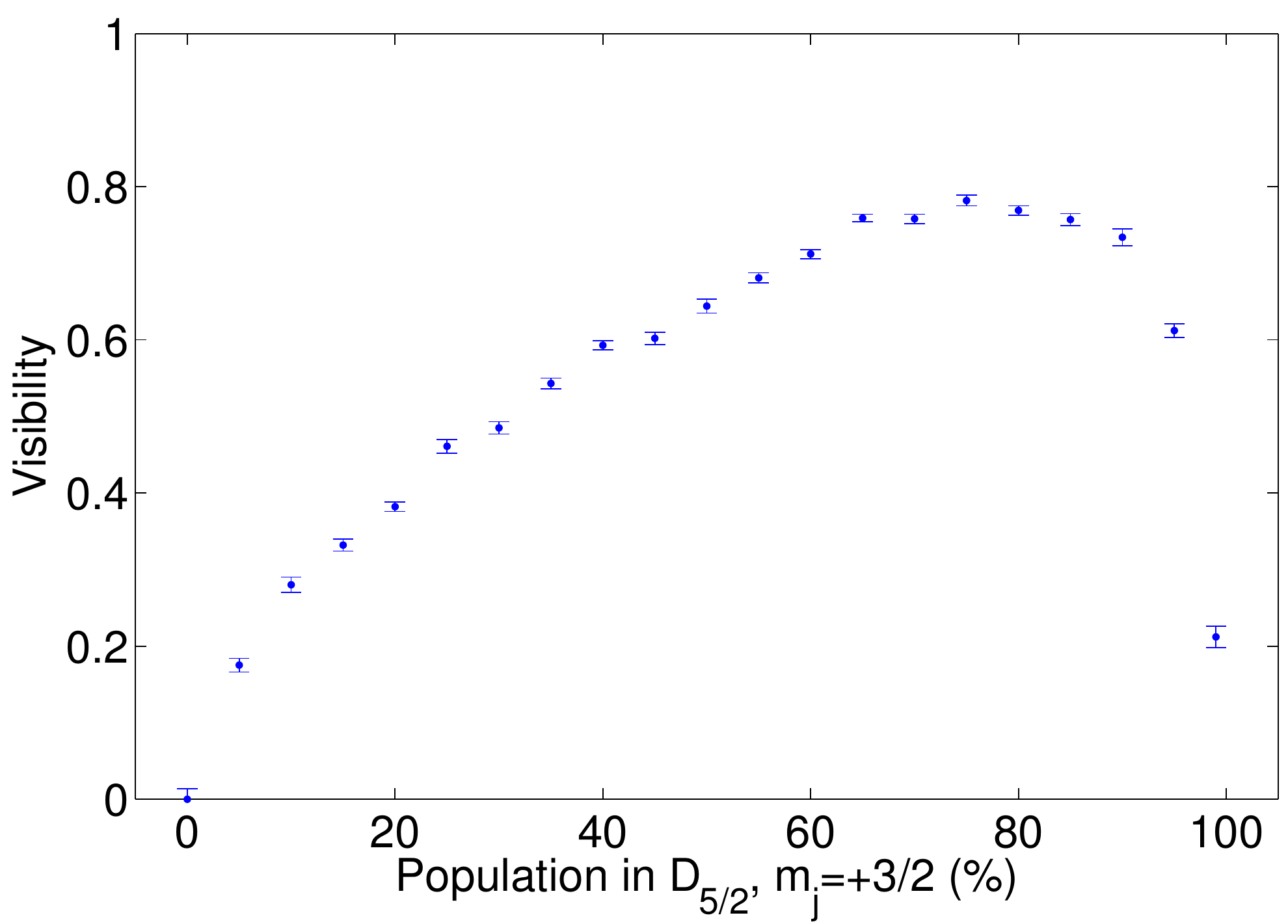}
\caption{(Color online) Quantum-beat visibility in the V scheme as a function of the initial population in $|\rm{D}_{5/2},+\frac{3}{2}\rangle$. The data points are calculated from exponential fits to the envelopes of the observed quantum beats.}
\label{best-pop}
\end{figure}

Due to the spurious decay from $|\rm{P}_{3/2},+\frac{1}{2}\rangle$ to $|\rm{S}_{1/2},+\frac{1}{2}\rangle$ [see Fig.~\ref{levelscheme_both}(b)] in the V scheme, maximum visibility of the quantum beats requires an unequal population ratio in the initial superposition state, as can be seen from Eq.~(\ref{V-superposition}). The data of Fig.~\ref{V-osc-pol+phase} are acquired with the optimum population ratio, which is experimentally verified by the following procedure: From the two pulses which prepare the coherent superposition, the duration of the first $\left\lvert\rm{S}_{1/2},-\frac{1}{2}\right\rangle \to \left\lvert\rm{D}_{5/2},+\frac{3}{2}\right\rangle$ pulse is varied, thereby adjusting the amount of transferred population. The subsequent $\pi$ pulse to $\left\lvert\rm{D}_{5/2},-\frac{5}{2}\right\rangle$ transfers the remaining population. In Fig.~\ref{best-pop} the quantum-beat visibility is shown as a function of the population in $\left\lvert\rm{D}_{5/2},+\frac{3}{2}\right\rangle$. The highest visibility of 78.2(9)$\%$ is obtained for 75$\%$ population, in agreement with the ratio of the two involved CGCs. 

The reduced visibility of the quantum beats in the V scheme (Fig.~\ref{V-osc-pol+phase}) compared to the $\Lambda$ scheme (Fig.~\ref{L-osc-pol+phase}) results from the nature of the respective interference phenomena. In the $\Lambda$ scheme the excitation amplitudes to the P level interfere constructively or destructively; hence, the overall probability for 393-nm photon emission is modulated. In contrast, in the V scheme the photons are emitted with a spatially rotating pattern. Therefore, the HALO that collects the 393-nm photons in $\sim 4\%$ solid angle will also pick up a small fraction of the dipolar emission when the preferred emission direction is perpendicular to the HALO axis.

\subsection{B. Phase-dependent photon-scattering probability}
\begin{figure}[htbp]
\centering
\includegraphics[width=8.3cm]{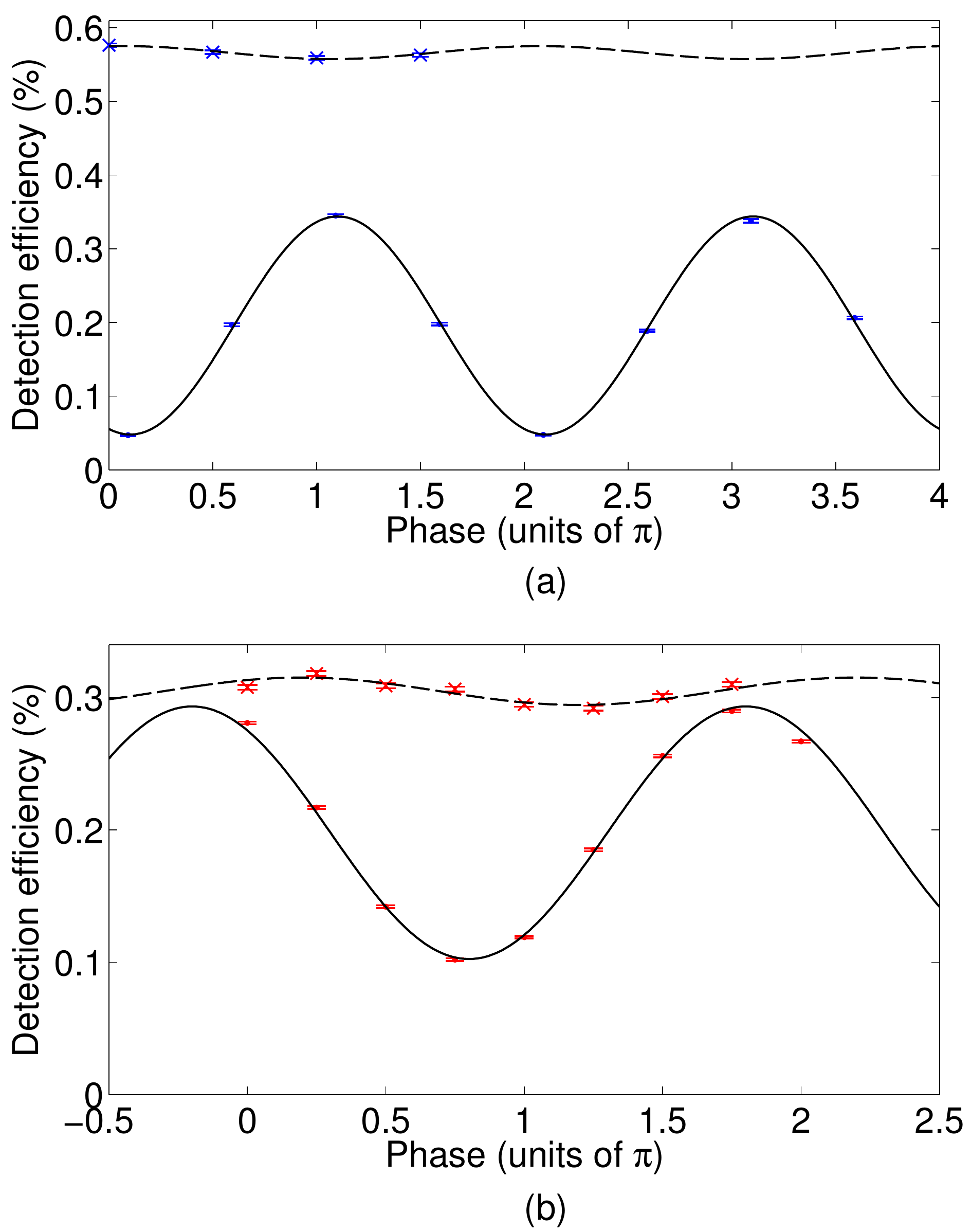}
\caption{(Color online) Suppression and enhancement of the generation efficiency of 393~nm photons controlled via the atomic phase $\Phi_{\rm D}(0)$. The dots and solid curves (sinusoidal fits) are for short 854~nm excitation pulses (duration $<$ quantum-beat period), while the crosses and dashed curves are for long pulses (duration $\gg$ quantum-beat period). (a) $\Lambda$ scheme: The short photon has about 70~ns duration and 553~ns quantum-beat period, the long photon about 850~ns duration and 106.5~ns quantum-beat period. (b) V scheme: The short photon extends over $\sim$120~ns at 277.4~ns beat period, for the long photon the values are 750 and 53.2~ns. }
\label{det-eff-both}
\end{figure}

The quantum beats in Figs.~\ref{L-osc-pol+phase} and ~\ref{V-osc-pol+phase} extend over many periods of the underlying Larmor precession; therefore, the total probability to detect a photon (i.e., time-integrated over the whole wave packet) is nearly independent of the control phases $\Phi_{854}$ and $\Phi_{\rm D}(0)$. This behavior changes, however, when the exciting laser pulse is shorter than the quantum-beat period (or at least of comparable duration). In this case the time-integrated probability of detecting a photon may be significantly enhanced or suppressed by the quantum interference. In order to emphasize this effect, we created quantum beats with high 854-nm laser power, such that the total duration of the generated photon wave packet was as short as $\sim$70~ns (including about 60~ns of broadening by the acousto-optic modulator rise time); at the same time, we reduced the magnetic field to extend the Larmor period. In Fig.~\ref{det-eff-both}, the time-integrated photon detection probability for the case of a short excitation pulse is plotted against the atomic control phase $\Phi_{\rm D}(0)$ (dots). With the $\Lambda$-type level scheme [Fig.~\ref{det-eff-both}(a)] we find a variation by a factor of 7 ($\sim76\%$ visibility), while for the V-scheme [Fig.~\ref{det-eff-both}(b)] the ratio is about 3 ($\sim48\%$ visibility). For comparison, the phase-dependent photon detection probability for a long photon, covering many quantum-beat periods, is also shown (crosses). Here the modulation is hardly visible; it disappears asymptotically. Hence, in the regime of short excitation pulses compared to the Larmor-precession period, the atomic control phase serves as a knob to determine the probability with which a photon is scattered into the detector. With faster modulation of the exciting laser than what could be attained in our setup, even much larger enhancement-suppression ratios may be reached. 

\section{V. PHYSICAL MECHANISM}

\begin{figure}[htbp]
\centering
\includegraphics[width=8.3cm]{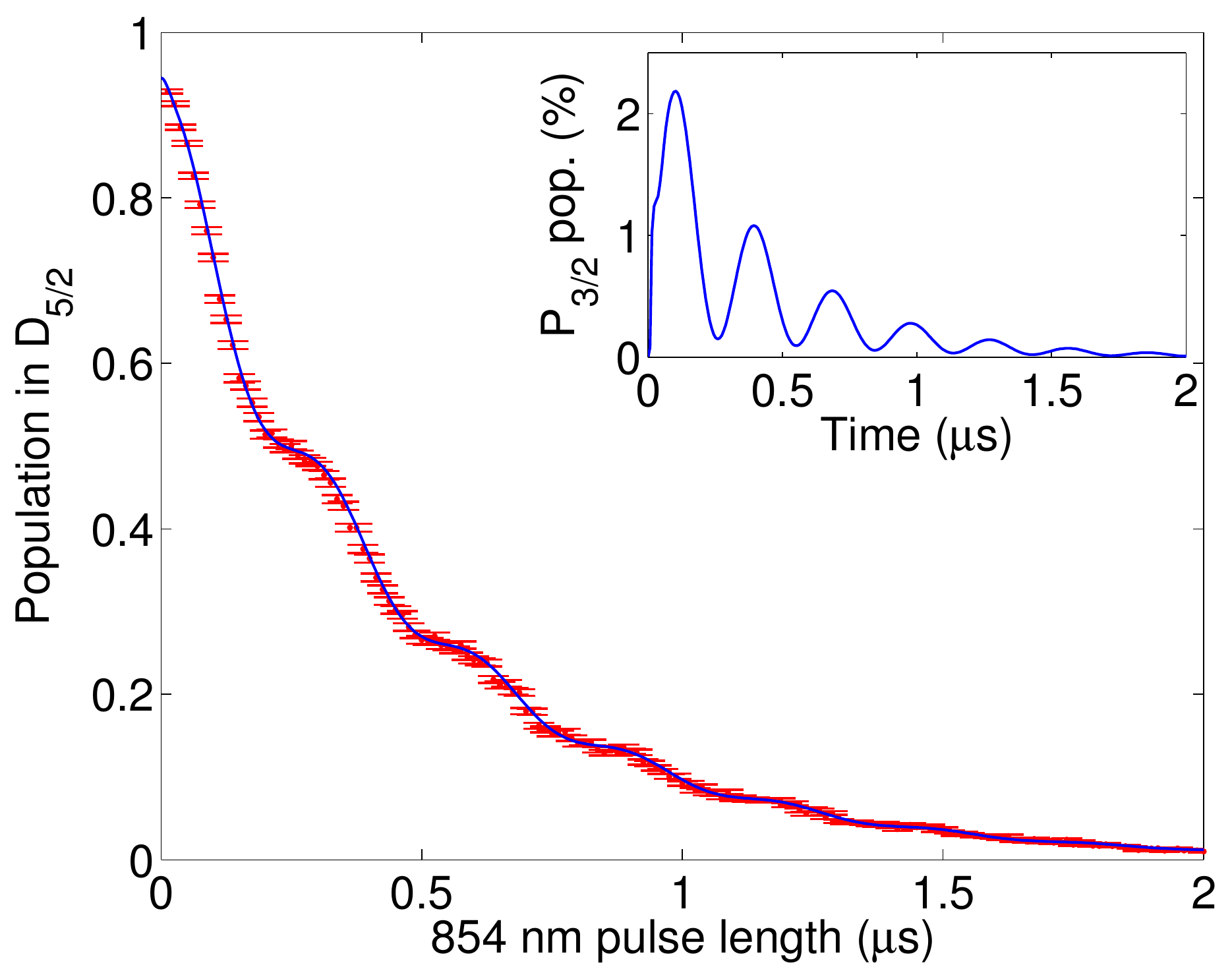}
\caption{(Color online) Change of the D$_{5/2}$ population for different 854-nm pulse lengths in the $\Lambda$ scheme. The stairs-like behavior results from interference of two excitation amplitudes to $\left\lvert\rm{P}_{3/2},-\frac{1}{2}\right\rangle$. The pulse length is varied in steps of 12.5~ns. The solid line is calculated with the 18-level Bloch equations including experimental parameters. (Inset) Time derivative of the calculated solid line showing the oscillation of the population in P$_{3/2}$, which leads to suppression and enhancement of the photon emission at 393~nm.}
\label{L-stairs}
\end{figure}

In this final section we highlight the fundamental difference between the interference processes involved in the two schemes. This is done by analyzing the time dependence of the population in D$_{5/2}$ during the excitation with 854~nm light: In the experiment, the excitation is interrupted after a certain pulse duration and the remaining population in the D level is determined through state-selective fluorescence \cite{Roos1999}. For the $\Lambda$ scheme, the result is displayed in Fig.~\ref{L-stairs}. Here the input polarization is set to $\left\lvert\rm{V}\right\rangle$, and the magnetic field has a value of 0.987~G, which results in a Larmor period between the Zeeman sublevels of 302 ns. The pulse length of the 854-nm excitation is varied in steps of 12.5 ns. The figure shows that the depopulation of the D$_{5/2}$ manifold exhibits a "stairs"-like modulation with the Larmor frequency. The population dynamics are very well fitted by 18-level Bloch equations (solid line in Fig.~\ref{L-stairs}). The inset in Fig.~$\ref{L-stairs}$ shows the time derivative of the population in D$_{5/2}$, which is proportional to the population in P$_{3/2}$. Comparison with Fig.~\ref{L-osc-pol+phase} confirms that the observed quantum beats in the case of the $\Lambda$ scheme are due to a corresponding oscillation of the population of the excited P level. This manifests that the interference happens indeed between the two absorption pathways from D$_{5/2}$ to P$_{3/2}$: As these require $\sigma^+$- and $\sigma^-$-polarized light, respectively, the Larmor-precessing phase of the initial superposition state creates an oscillatory excitation probability for any fixed linear incoming polarization.

\begin{figure}[htbp]
\centering
\includegraphics[width=8.3cm]{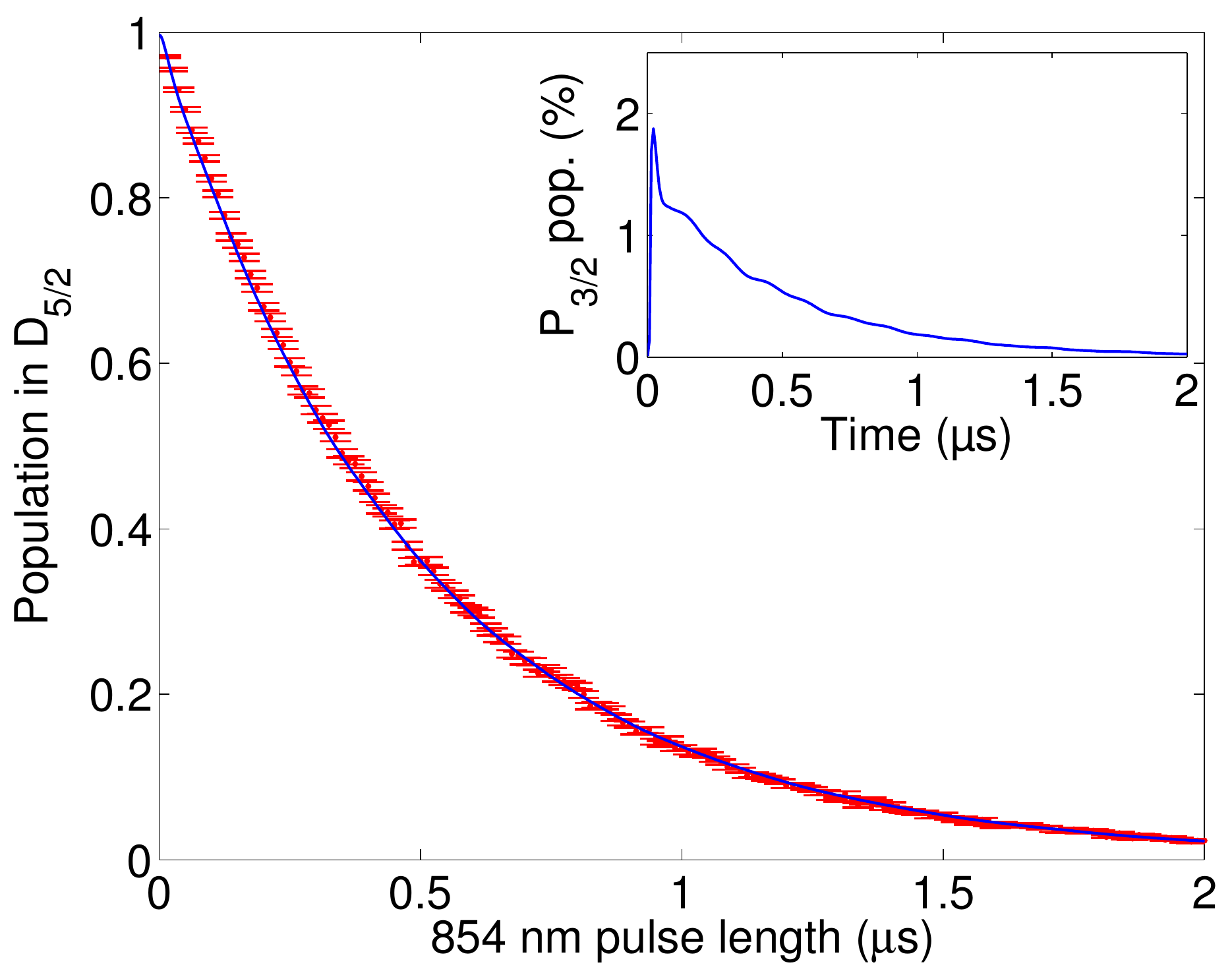}
\caption{(Color online) Depopulation of D$_{5/2}$ for different 854-nm pulse lengths in the V scheme, showing simple exponential decay without modulation. The points are measured data; the line is a fit by Bloch equation dynamics. (Inset) Time derivative of the calculated curve \cite{footnoteV}.}
\label{V-nostairs}
\end{figure}

In contrast, a similar measurement for the case of the V scheme, shown in Fig.~\ref{V-nostairs}, does not exhibit any modulation on the exponential depopulation curve of the D$_{5/2}$ level. The Bloch equation fit describes excitation from D$_{5/2}$ to P$_{3/2}$ at a constant rate, i.e., with no interference in the excitation pathways. The observed quantum beats in the emitted 393-nm photon wave packet are therefore due to interference of emission amplitudes: The Larmor precession of the initial state enters into the emitted superposition of $\sigma^+$- and $\sigma^-$-polarized components and causes a spatially rotating emission pattern, which leads to an oscillatory detection probability. 

\begin{figure}[htbp]
\centering
\includegraphics[width=8.3cm]{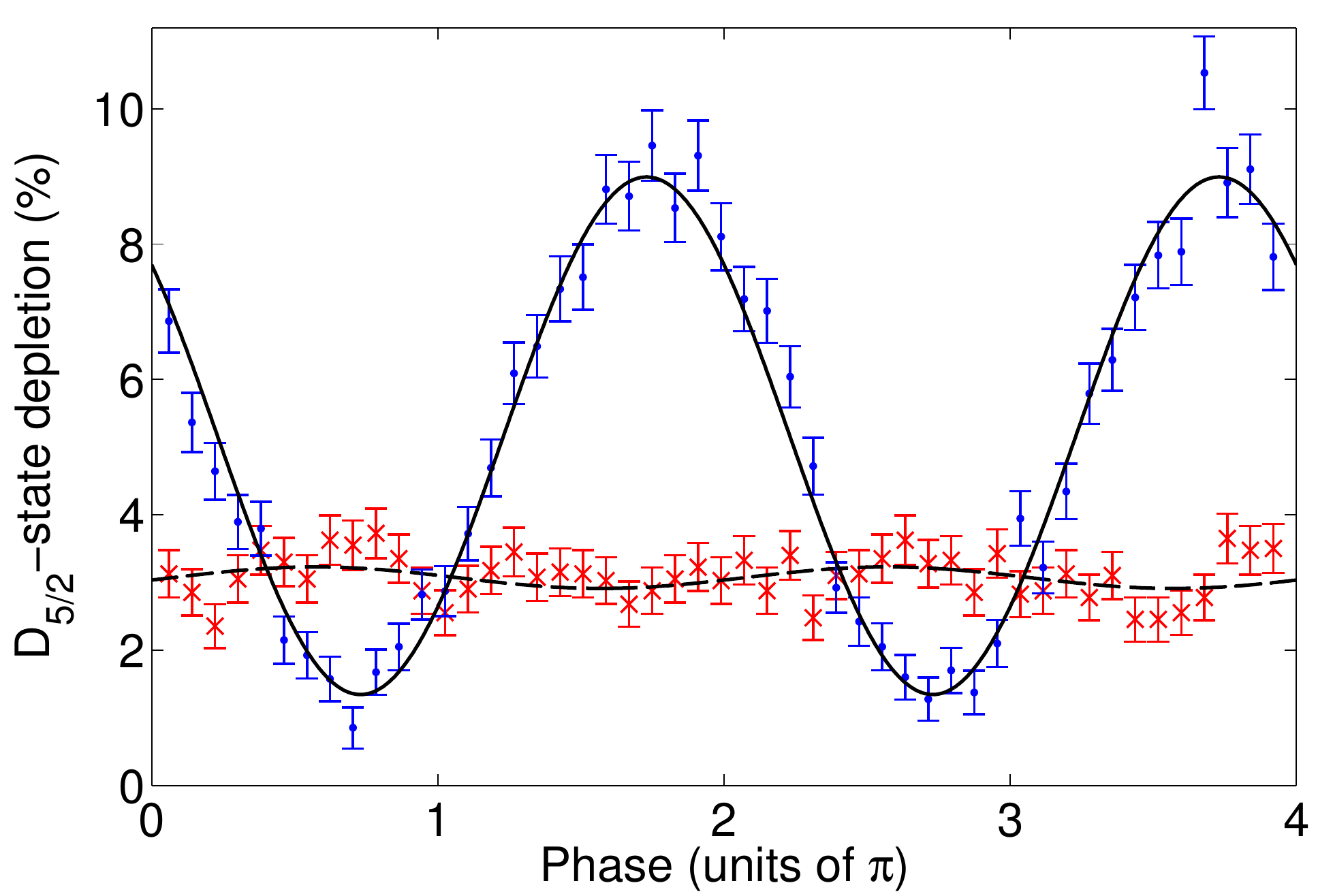}
\caption{(Color online) Depopulation of D$_{5/2}$ for different intial phases $\Phi_{\rm D}(0)$ of the initial superposition, after an excitation pulse of 12.5~ns. The blue dots (solid curve) are measured (calculated) for the $\Lambda$ scheme, the red crosses (dashed curve) for the V scheme. The Larmor period was 300~ns ($\Lambda$) and 150~ns (V).}
\label{depletion}
\end{figure}

Further evidence for our explanation of the physical mechanism is provided by measuring the depletion of the D$_{5/2}$ level as a function of the initial phase, $\Phi_{\rm D}(0)$, after an excitation pulse of fixed duration. A comparison of the results for the two distinct level schemes, under otherwise equal conditions, is displayed in Fig.~\ref{depletion}. The phase enters into the depletion in the $\Lambda$ case, while the V case is practically insensitive to it. 
 
The two types of quantum beats of Figs.~\ref{L-osc-pol+phase} and \ref{V-osc-pol+phase}, while looking very similar, are hence identified to have manifestly different physical origins. 

\section{V. Summary}
We investigated experimentally two cases of single-photon quantum beats originating from interference of photon-scattering amplitudes. A single trapped Ca$^+$ ion is prepared in a coherent superposition state in its D$_{5/2}$ manifold and then excited on the D$_{5/2}$ to P$_{3/2}$ transition (854~nm), whereby it releases a single photon on the P$_{3/2}$ to S$_{1/2}$ transition (393~nm). Quantum beats with the frequency of the atomic Larmor precession are manifested as oscillations in the photon arrival-time distribution with $>93\%$ visibility. The phase of the quantum beats is controlled through setting the phase of the initial atomic superposition and through the polarization of the exciting 854-nm light. For a $\Lambda$- and a V-type atomic level scheme, we identified interference of absorption and emission amplitudes, respectively, to be the physical mechanisms behind the quantum beats. The two mechanisms are fundamentally different in that the remaining population in the D$_{5/2}$ level reveals the interference in the $\Lambda$ case but not in the V case. As a consequence, for small ratios between the excitation pulse length and the Larmor period, excitation out of D$_{5/2}$ can be efficiently suppressed and enhanced in the $\Lambda$ scheme through the choice of the control phases, while in the V case the excitation probability is phase insensitive. The presented experimental techniques and the physical mechanisms are essential ingredients for mapping arbitrary polarization states of photons onto a single ion \cite{Kurz2014}.

\vspace{0.2cm}
\begin{acknowledgments}
We acknowledge support by the BMBF (Verbundprojekt QuOReP, CHIST-ERA project QScale), the German Scholars Organization / Alfried Krupp von Bohlen und Halbach-Stiftung, and the ESF (IOTA COST Action).
\end{acknowledgments}

{}


\begin{thebibliography}{99}
\bibitem{footnoteCGC}
The definition that we use for the CGCs is related to ''standard'' notations through $C_{m_{\text{D}},m_{854},m_{\text{P}}}=C^{5/2,1,3/2}_{m_{\text{D}},m_{854},m_{\text{P}}}=\langle5/2,m_{\text{D}};
  1,m_{854}|5/2,1,3/2,m_{\text{P}}\rangle$, and likewise for
  $C_{m_{\text{P}},m_{393},m_{\text{S}}}$.
\bibitem{footnoteV}
In contrast to the data of Fig.~\ref{V-osc-pol+phase}, the data of this figure
  was recorded with an initial atomic superposition of equal amplitudes. This
  does not affect our conclusions.
\end{thebibliography}
\end{document}